\documentclass[twocolumn]{raa}
\usepackage{graphicx,times}             
\usepackage{natbib}
\usepackage{amssymb,amsmath}
\usepackage{multirow}
\usepackage{threeparttable}
\bibpunct{(}{)}{;}{a}{}{,}

\usepackage[pagebackref=true]{hyperref}

\begin{document}
	
	\title{PyMsOfa: A Python Package for the Standards of Fundamental Astronomy (SOFA) Service
	}

	\volnopage{Vol.0 (20xx) No.0, 000--000}      
	\setcounter{page}{1}          
	
	\author{ Jianghui Ji$^{1,2,3}$
		\and Dongjie Tan$^{1,2}$
		\and Chunhui Bao$^{1,2}$
		\and Xiumin Huang$^{1,2}$
		\and Shoucun Hu$^{1,2,3}$
		\and Yao Dong$^{1}$
		\and Su Wang$^{1,3}$
	}
	
	\institute{CAS Key Laboratory of Planetary Sciences, Purple Mountain Observatory, Chinese Academy of Sciences,
		Nanjing 210023, China; {\it jijh@pmo.ac.cn}\\
		\and
		School of Astronomy and Space Science, University of Science and Technology of China, Hefei 230026, China\\
       \and
       CAS Center for Excellence in Comparative Planetology, Hefei 230026, China\\
	\vs\no
	{\small Received 20xx month day; accepted 20xx month day}}
	
\abstract{The Standards of Fundamental Astronomy (SOFA) is a service provided by the International Astronomical Union (IAU) that offers algorithms and software for astronomical calculations, which was released in two versions by FORTRAN 77 and ANSI C, respectively. In this work, we implement the python package PyMsOfa for SOFA service by three ways: (1) a python wrapper package based on a foreign function library for Python (ctypes), (2) a python wrapper package with the foreign function interface for Python calling C code (cffi), and (3) a python package directly written in pure python codes from SOFA subroutines. The package PyMsOfa has fully implemented 247 functions of the original SOFA routines. In addition, PyMsOfa is also extensively examined, which is exactly consistent with those test examples given by the original SOFA. This python package can be suitable to not only the astrometric detection of habitable planets of the Closeby Habitable Exoplanet Survey (CHES) mission \citep{ji2022ches}, but also for the frontiers themes of black holes and dark matter related to astrometric calculations and other fields. The source codes are available via https://github.com/CHES2023/PyMsOfa.
	}
	
	\authorrunning{Jianghui Ji,  Dongjie Tan,  Chunhui Bao, et al.}            
	\titlerunning{PyMsOfa: A Python Package for the SOFA Service}  
	
	\maketitle

\section{Introduction}
\label{sect:intro}
	
The Standards of Fundamental Astronomy (SOFA)\footnote{http://www.iausofa.org/index.html} service was established by the International Astronomical Union (IAU), which is an algorithm and program based on the fundamental astronomical models that aims to show the IAU resolutions in an authoritative program. The SOFA service contains the implementation of the time system, the transformation of the coordinate system, and the attitude of the earth (e.g., the precession and nutation, closely related to International Earth Rotation Service), and astrometric parameters. The routines were well written in strict compliance with IAU resolutions and have been updated as the resolutions are amended, with the latest version released on 2021 May 12.
	
The SOFA service is officially released in two programming languages, i.e., FORTRAN 77 and ANSI C, whose advantages of high efficiency and fast running make them widely utilized in scientific computations. However, the python language greatly benefits from its easy-to-learn, simple syntax, and tremendous built-in libraries in science and engineering, which motivates us to implement the full python package {PyMsOfa\footnote{https://github.com/CHES2023/PyMsOfa}} for SOFA, leading to a direct, convenient and efficient routines available.
	
There are several python packages that contain SOFA features, such as astropy\footnote{https://docs.astropy.org/en/stable/index.html},  PyERFA\footnote{https://github.com/liberfa/pyerfa, https://doi.org/10.5281/zenodo.3940699}, {pysofa\footnote{https://pypi.org/project/pysofa} (which contains only 186 subroutines from 247 SOFA services)}. However, in this work, the python package is a complete implementation of all 247 functions of a recently released SOFA version in three ways via python, i.e., wrapped with the ctypes library and the cffi interface, and written directly in pure python codes based on SOFA subroutines. Thus the users of SOFA with FORTRAN 77 and ANSI C version can easily get started with this python package, and achieve similar goals in the study.
	
The paper is organized as follows: in Section ~\ref{sect:Functions} we briefly describe several major modules that SOFA services can achieve. Section ~\ref{sect:python} describes the features of python package and presents two examples of astrometric calculations. Finally, we make a concise summary.	
	
	\section{Functions in SOFA service}
	\label{sect:Functions}
	
The routines of the SOFA service can be divided into four categories: basic calculation module, time scale and calendar module, coordinate system transformation module, earth attitude module. Each consists of dozens of routines. A brief introduction for these routines and functions are described below. More detailed descriptions can be found in the comments and SOFA documentation.
	
	\subsection{Basic calculation module}
Such routines are very basic, mainly involving the processing of parameters, vectors and matrices. Its purpose is to conduct the calling between SOFA routines, so it does not fully include all vector and matrix operations. Similar functionality can be implemented in many other packages.
	
	These routines can realize the conversion of coordinates, such as the conversion between spherical coordinates to Cartesian (e.g. pymS2c and pymC2s). The routines about vector contains the processing of velocity - position vectors that are required in other programs (e.g. pymTrxpv and pymPvu). For matrix, it contains functions such as rotation of matrix (e.g. pymRx). In addition, the SOFA service also provides procedures for projective relations (e.g. pymTpxes), involving the conversion between spheres and planes.
	
	\subsection{Time scale and calendar module}
	
\begin{figure}
		\centering
		\includegraphics[width=80mm, angle=0]{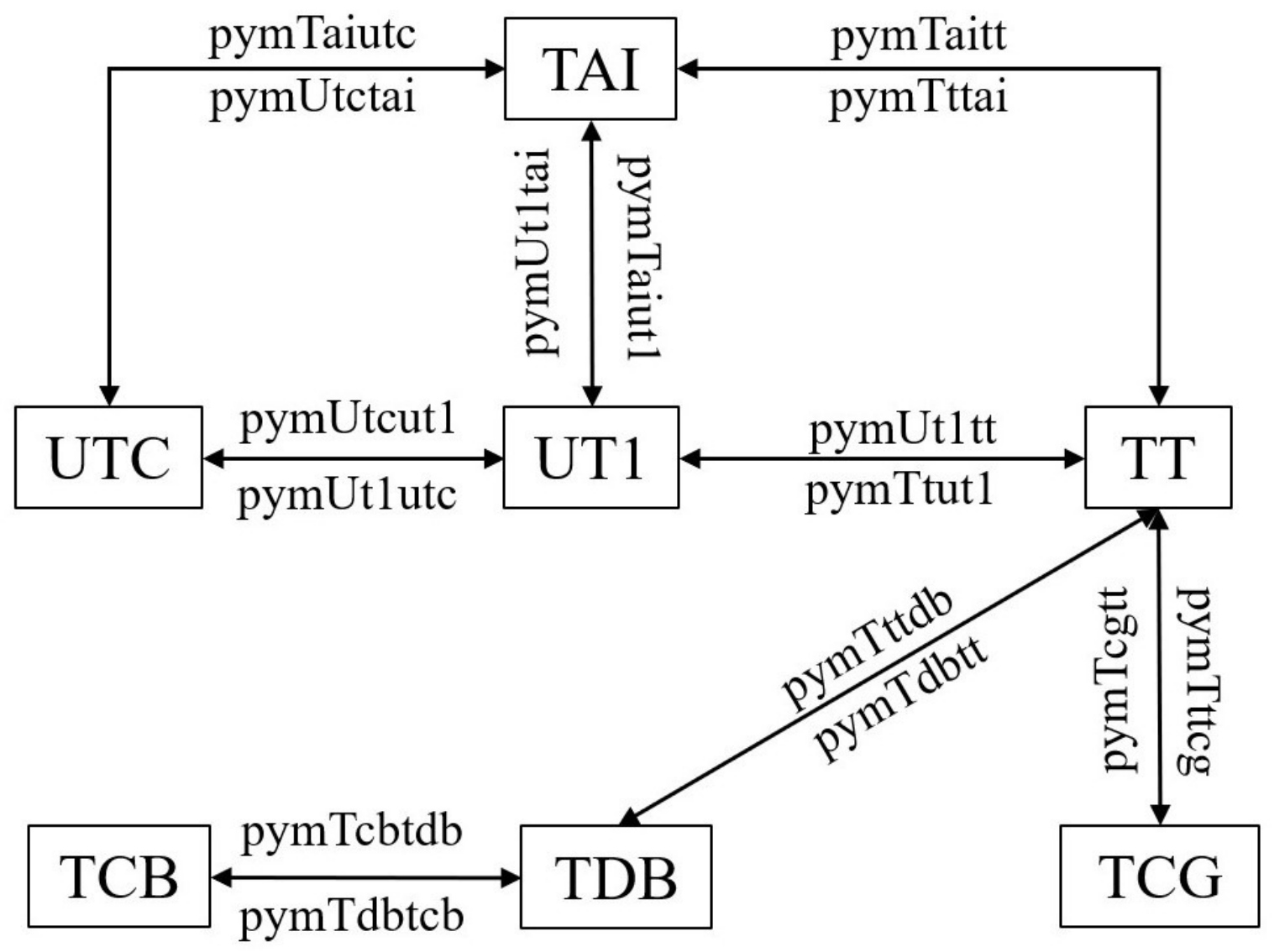}
		\caption{The functions involved in the conversion between the time scales in Table~\ref{Tab1}. }
		\label{Fig1}
	\end{figure}

The time scales involved in the SOFA services are shown in Table~\ref{Tab1}. Each time scale has a different use, and there is a slight time difference between them \citep{moyer1981transformation,fairhead1990analytical,muller2012transactions}. Some of the time scales have a linear relationship, while others have to be transformed according to the location \citep{seidelmann1992explanatory,soffel2003iau,mccarthy2004iers}. Figure~\ref{Fig1} shows the functions involved in the conversion between time scales. In addition, the discontinuity of time caused by leap seconds also needs to be taken into account (e.g. pymDat) \citep{seidelmann1992explanatory}. Therefore, when performing astronomical scientific calculations, it is necessary to choose an appropriate time scale.

	\begin{table}
	\begin{center}	
		\caption[]{The time scales involved in SOFA services.}\label{Tab1}
		\begin{tabular}{c|c|l}
			\hline
			\hline
			{Name}          	 		  & {Abb.}       & {Description}                       								\\ \hline
			International Atomic Time         & TAI                 & Widely used time for the concrete realization of TT                      \\
			Coordinated Universal Time        & UTC                 & The universal standard time                                              \\
			Universal Time                    & UT1                 & Determined by the rotation of the Earth                                  \\
			\multirow{2}{*}{Terrestrial Time} & \multirow{2}{*}{TT} & Coordinate time, can be used as the time argument of geocentric calendar \\
			&                     & former name: TDT                                                         \\
			Geocentric Coordinate Time        & TCG                 & Coordinate time for the Geocentric Celestial Reference System (GCRS)     \\
			Barycentric Coordinate Time       & TCB                 & Coordinate time for the Barycentric Celestial Reference System (BCRS)    \\
			Barycentric Dynamical Time        & TDB                 & Time variable of the current solar system planetary calendar             \\ \hline
			\hline
		\end{tabular}

	\end{center}
	\end{table}
	
The Julian Date is an astronomical method that measures time over a long time span, which is widely adopted in the expression of calendar. In the SOFA service, the Julian Date is described by two parameters,  for a specific calendar of Julian Day JD = 2450123.7, the date can be denoted as JD1 = 2400000.5 and JD2 = 50123.2, respectively.

\subsection{Coordinate system transformation module}
The SOFA services include the conversion of different coordinate systems, such as geodetic coordinate, geocentric coordinate, horizon coordinate, hour angle system of coordinate, ICRS, ecliptic coordinate and galactic coordinate. The package provides a complete system for the astrometric parameters, 'ASTROM', which is composed of 30 components. These components include the time, the solar system barycenter (SSB) to the observer, the direction and distance relative to the Sun, the barycentric observer velocity, reciprocal of Lorenz factor, bias-precession-nutation matrix, TIO locator \citep{mccarthy2004iers}, polar motion, geodetic latitude, diurnal aberration \citep{klioner2003practical}, Earth rotation angle and refraction constant A, B \citep{crane1976refraction, rueger2002refractive}. Detailed calculation procedures are also provided for each component. The final ASTROM parameters are the astrometric parameters of different stations, independent of stars. Table~\ref{Tab2} shows several major basic coordinate systems involved in SOFA services. Here Table~\ref{Tab3} shows the functions that produce the components of the ASTROM and the functions used to transform the stellar parameters.
	
It also deals with the transitions between FK4, FK5 and the Hipparcos catalog (e.g. pymFk425, pymFk524 and pymFk5hip)  \citep{aoki1983conversion,yallop1989mean,mignard2000global}. These coordinate systems cover the fundamental coordinate systems in classical astrometry. In astronomical calculation, the change of coordinate system is inseparable from the change of position of the observation target to various observation stations \citep{smart1962spherical}. The SOFA service here gives a set of authoritative algorithms.

\begin{table}
\begin{center}
		\caption[]{The fundamental coordinate systems involved in SOFA services.}\label{Tab2}

		\begin{tabular}{c|c|l}
			\hline
			\hline
			{Name}           & {Abb.}         & {Description}                       					\\ \hline
			International Celestial & \multirow{2}{*}{ICRS} & An ideal coordinate system that does not move with 			\\
			Reference System        &                       &  respect to distant extragalactic objects.                   	\\ \hline
			Celestial Intermediate  & \multirow{2}{*}{CIRS} & Geocentric reference system modified by precession       		\\
			Reference System        &                       & and nutation based on GCRS.                                	\\ \hline
			Geocentric Celestial    & \multirow{2}{*}{GCRS} & Geocentric space-time coordinate system within the    		\\
			Reference System        &                       & framework of general relativity.                   			\\ \hline
			Barycentric Celestial   & \multirow{2}{*}{BCRS} & A set of space-time coordinate systems of the               	\\
			Reference System        &                       & solar system centroid.                                        \\ \hline
			\hline
		\end{tabular}
	
	\end{center}
\end{table}
	
	\begin{table}
	\begin{center}
		\caption[]{The transformation between basic coordinate systems.}\label{Tab3}
		\begin{threeparttable}
		\begin{tabular}{c|c|c|c}
			\hline
			\hline
			{Observer}             & {Transformation} & {Parameters produce\tnote{1}} & {Application\tnote{2}}                                                     \\ \hline
			space                        & ICRS $\leftrightarrow$ GCRS      & pymApcs pymApcs13  & \multirow{3}{*}{pymAtciq* pymAticq*}                                            \\ \cline{1-3}
			geocentric                   & ICRS $\leftrightarrow$ GCRS      & pymApcg pymApcg13  &                                                                                 \\ \cline{1-3}
			\multirow{3}{*}{terrestrial} & ICRS $\leftrightarrow$ CIRS      & pymApci pymApci13  &                                                                                 \\ \cline{2-4}
										& ICRS $\leftrightarrow$ Observed  & pymApco pymApco13  & \begin{tabular}[c]{@{}c@{}}pymAtciq* pymAticq*\\ pymAtioq pymAtoiq\end{tabular} \\ \cline{2-4}
										& CIRS $\leftrightarrow$ Observed  & pymApio pymApio13  & pymAtioq pymAtoiq                                                               \\ \hline
		\end{tabular}
		\begin{tablenotes}
		\footnotesize
		\item[1] The number 13 at the end of the functions indicates the use of internal functions to compute various ephemeris.
		\item[2] The symbol (*) indicates that the function has multiple variations, such as pymAtciqn, pymAtciqz.
		\end{tablenotes}
		\end{threeparttable}
	\end{center}
	\end{table}

In addition to this, the SOFA services still provides the transformation of star positions between FK4, FK5 and the Hipparcos catalogue.
	
	\subsection{Earth attitude module}
	
The SOFA service provides an authoritative definition of the parameters involved in the earth attitude and provides detailed algorithms. These parameters include the parameters of the objects in the solar system (e.g., the mean longitude of the planets and their approximate heliocentric positions and velocities, e.g. pymPlan94)  \citep{simon1994numerical,klioner2003practical}, bias-precession-nutation matrix \citep{lieske1979precession,mathews2002modeling,wallace2006precession}, Earth rotation angle \citep{capitaine2000definition}, the celestial intermediate pole (CIP) and celestial intermediate origin (CIO) \citep{capitaine2003expressions2,capitaine2003expressions,capitaine2006high}, etc (e.g. pymPn00 and pymNum06a). There are some differences in these parameters under different models, so SOFA also gives the parameters under the IAU 1976 model \citep{lieske1977expressions}, IAU 1980 model \citep{seidelmann19821980}, IAU 2000 A\&B model \citep{soffel2003iau}, and IAU 2006 model \citep{seidelmann2007report}. The calculations of the routines are also consistent with those in the IAU resolution \citep{wallace2006precession}. Thus one can choose the required models and parameters with respect to different astronomical calculations.
	
\section{Python feature of the package}
\label{sect:python}
For simplicity, here this python package of SOFA service is described based on the ctypes library, which is a foreign function library for Python that provides C compatible data types, and allows calling functions in the shared libraries\footnote{The cffi interface version and pure python codes are also available via the GitHub website.}. We integrated all 247 functions in a C file from SOFA package, which is named 'sofa\_a.c'. A dynamic shared library file (libsofa\_c.so) was compiled from three source files (sofa.h, sofam.h and sofa\_a.c). By calling the shared library file and the ctypes library, we establish a python interface for each SOFA function under ANSI C, thus SOFA services can be employed directly in python programs. The codes can be accessible at \href{https://github.com/CHES2023/PyMsOfa}{https://github.com/CHES2023/PyMsOfa}.

This python package differs from the ANSI C version in that the parameters involved in its functions vary in addition to the language used. In the original ANSI C version by SOFA, the input and output parameters of the function are included in the parameters of the calling function. In the Python version, the parameters of the function refer only to those input parameters. In addition, the parameters involved in each function, along with their types, units, etc., are briefly given in their comments. It can be easily queried when using each function by the package in the python environment. In the following, we will show two applications for the usage of the python package for SOFA.
	
\begin{figure}
		\centering
		\includegraphics[width=140mm, angle=0]{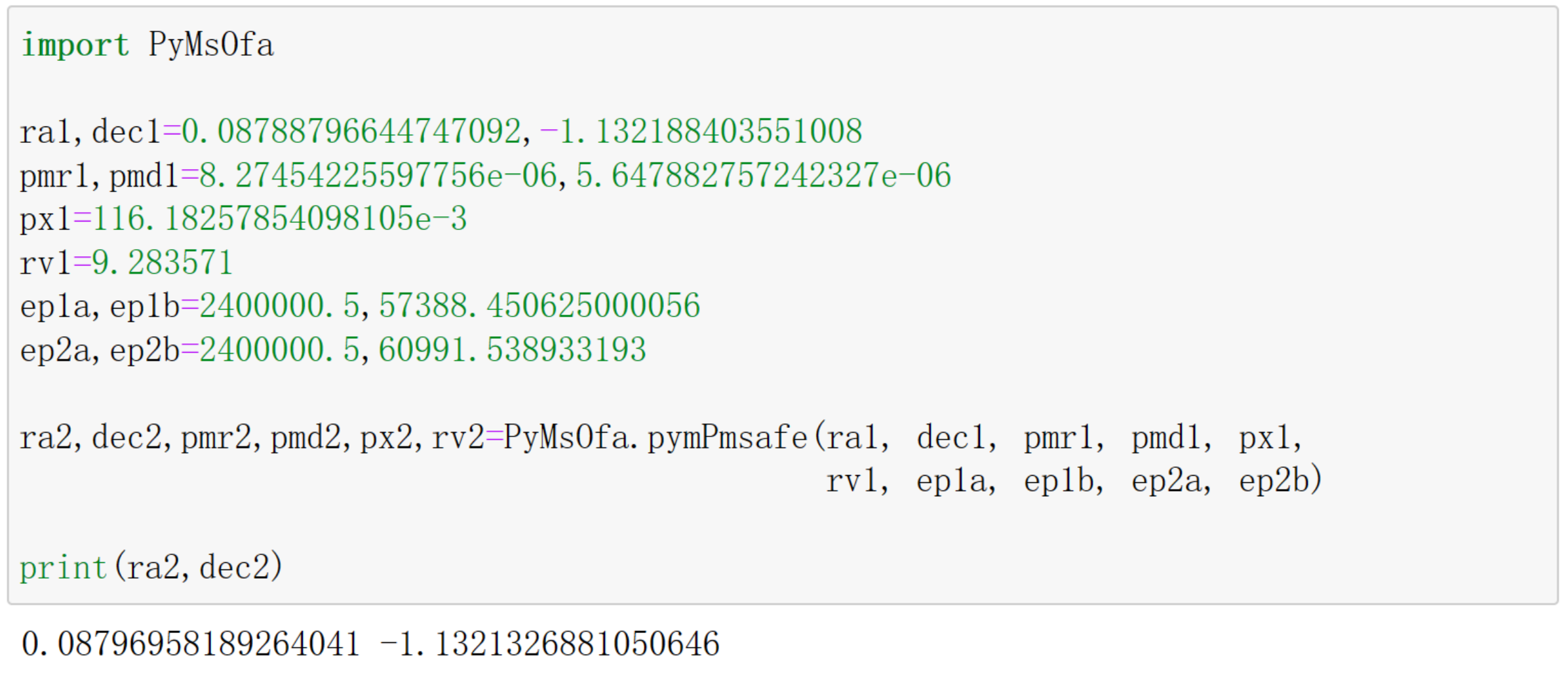}
		\caption{pymPmsafe: Example for the calculation of the RA and DEC at the given epoch.}
		\label{Fig2}
	\end{figure}

\begin{figure}
		\centering
		\includegraphics[width=140mm, angle=0]{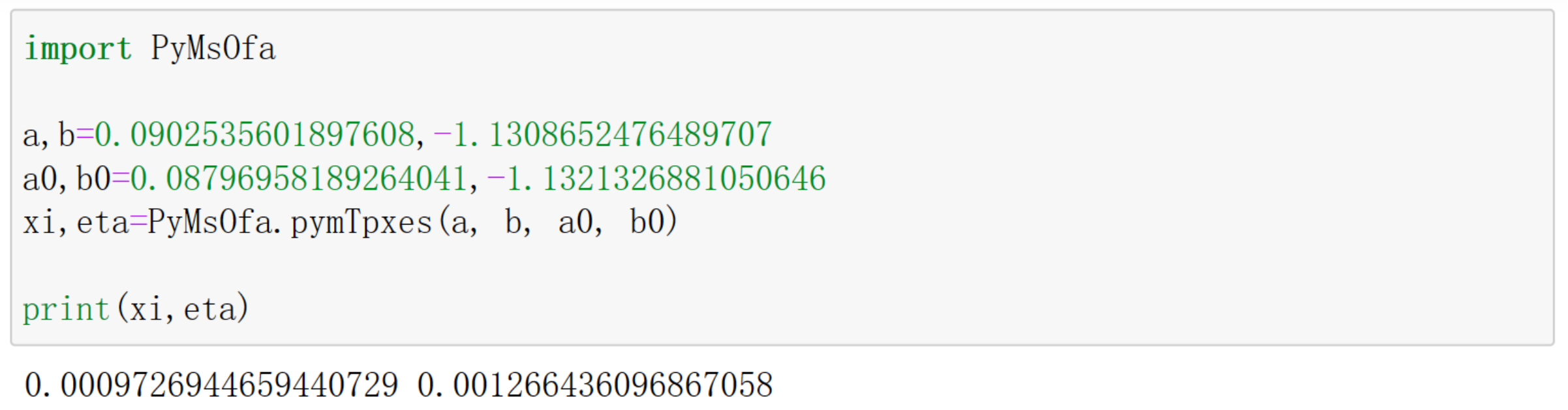}
		\caption{pymTpxes: Example for the calculation of the rectangular coordinates on focal plane.}
		\label{Fig3}
	\end{figure}

\begin{figure}
		\centering
		\includegraphics[width=\textwidth, angle=0]{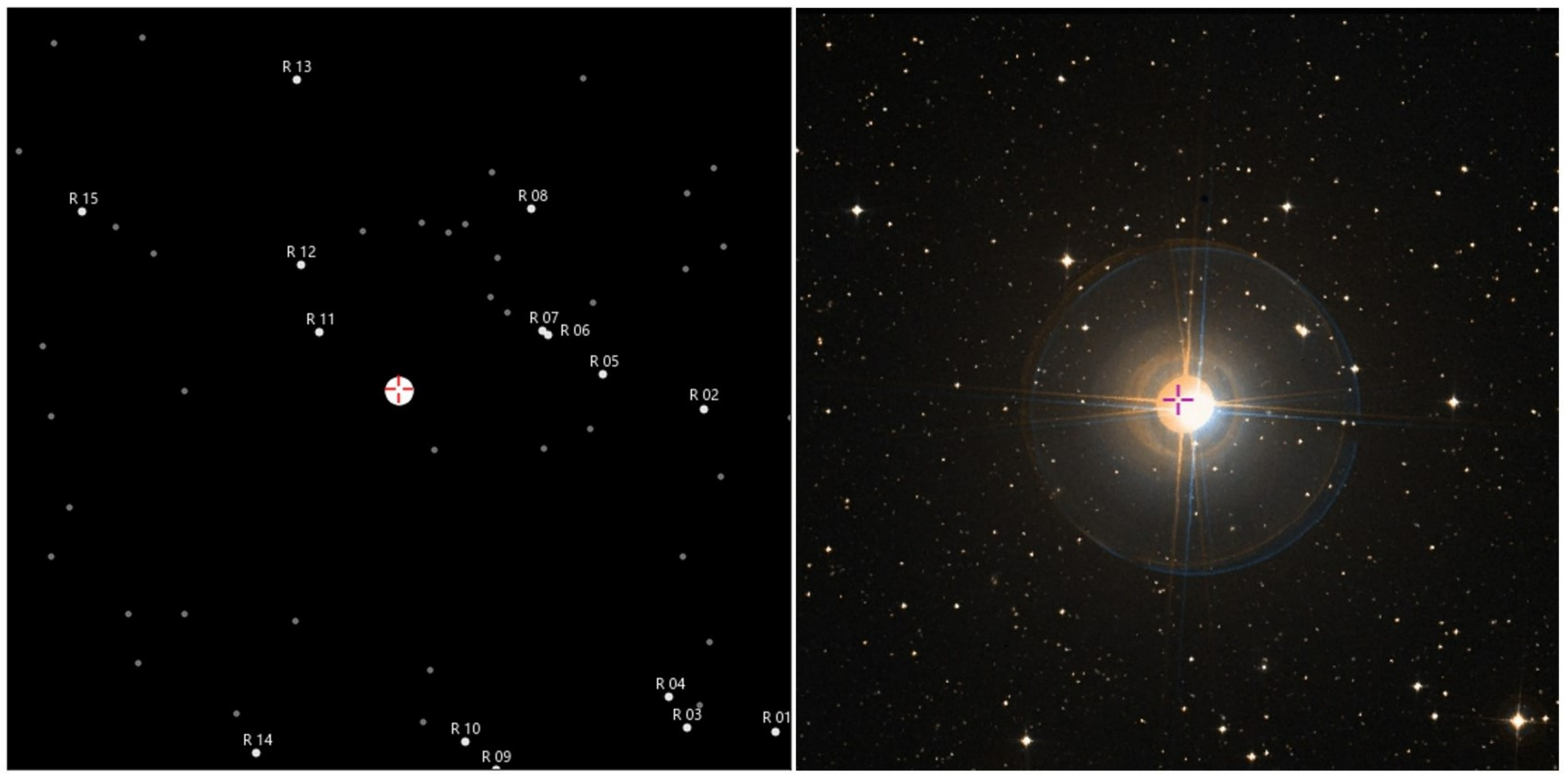}
		\caption{\textit{Left panel} : The simulation image of the target star (* zet Tuc) in CHES mission. The positions of the stars are based on the data of Gaia DR3 and converted to the predetermined observation time (JD2460992.04) and observation position (L2) by the relevant function in PyMsOfa. \textit{Right panel} : The image of the same field of view obtained from ESASky.}
		\label{Fig4}
	\end{figure}

For CHES, a number of functions in SOFA service are employed to simulate the observed images. Here we present two examples (see Fig.~\ref{Fig2} and Fig.~\ref{Fig3}). The first example illustrates the time expression in SOFA services as mentioned previously that the time is usually expressed by two parameters. The given two examples can accurately calculate the coordinates of the target star and reference stars based on the GAIA DR3, respectively, at the time of observation (on the plane). From the projection theorem, the coordinates in the focal plane are given to calculate the major observed quantity in CHES mission, i.e., the angular distance between them in the focal plane. Such functions can further provide references to relevant parameters in the CHES mission \citep{ji2022ches,tan2022evaluate}. Figure~\ref{Fig4} shows a simulation image with the above-mentioned two functions (pymPmsafe, pymTpxes), which is in good agreement with the produced image from ESASky.

	
\begin{figure}[!t]
		\centering
		\includegraphics[width=\textwidth, angle=0]{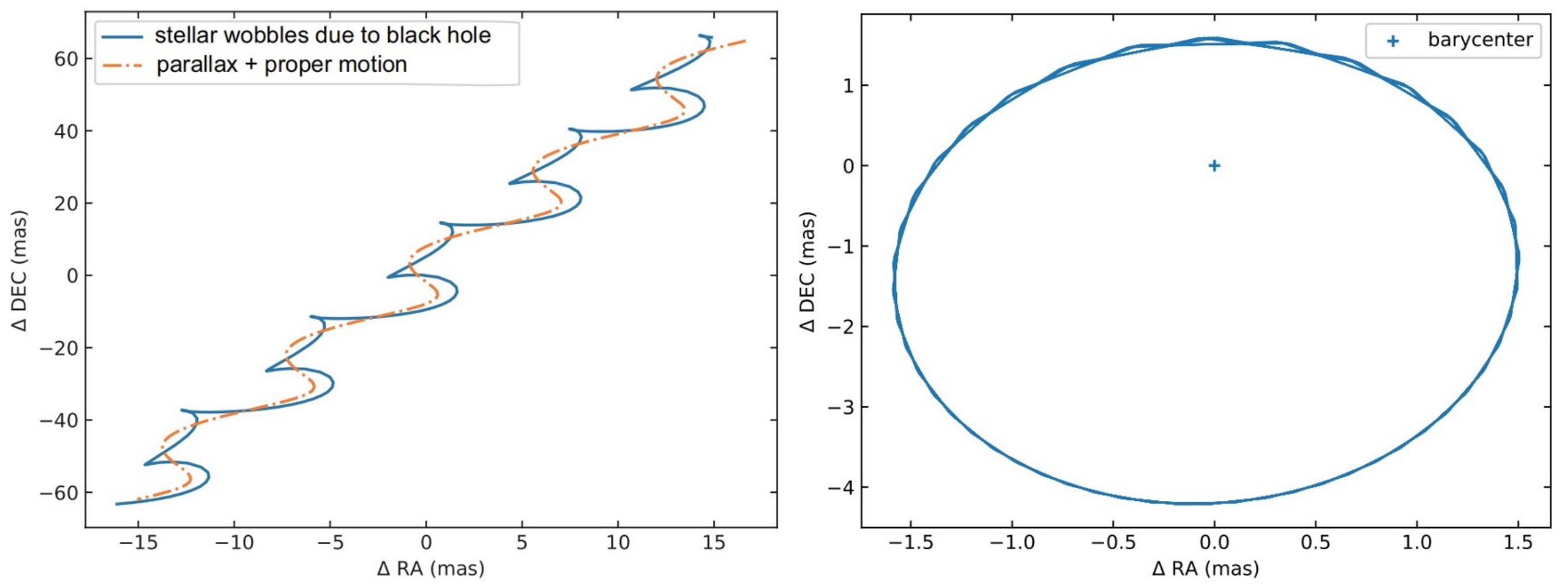}
		\caption{The predicted motion of Gaia BH1's photocenter on the sky during 5 years. \textit{Left panel} : The total motion due to proper motion, parallax, and orbital motion. The dash-dotted isolates the contribution due to parallax and proper motion alone. \textit{Right panel} : The predicted orbital motion, with parallax and proper motion removed.}
		\label{Fig5}
	\end{figure}

In addition to the detection of exoplanets, this package is also available for various astrometric observations, such as the detection of black holes, dark matter, etc. Fig~\ref{Fig5} reproduces the prediction of the motion of the star orbiting a black hole using PyMsOfa. The orbital parameters in the simulations are adopted from \citet{el2023sun}. Using this python package, the proper motion and parallax of the star can be carefully processed to achieve an accurate orbital motion under the gravitational from the black hole (Fig~\ref{Fig5}). This will contribute to a more detailed investigation and analysis of the physical properties of black holes.

\section{Summary}
\label{sect:summary}
{In this work we fulfill the python package PyMsOfa for SOFA services through three ways: (1) a python wrapper package based on ctypes library, (2) a python wrapper package with the cffi interface, and (3) a python package directly written in pure codes from SOFA algorithms. They implement SOFA services under python interface in various ways but each package can have the same usage.} The purpose of this work is to enable a wide variety of the usage of the authoritative algorithms in SOFA service under python interface. The python package has fulfilled all 247 functions of SOFA services written ANSI C, containing basic calculation, time scale and calendar, coordinate system transformation, earth attitude and astrometric parameters. This python package is thoroughly and correctly tested with those original examples given by SOFA, which runs stably on Linux, macOS and Windows operating systems.
	
Astrometry is an ancient branch of astronomy, which primarily aims at the investigation of the positions of celestial bodies. With quick development,  the astrometry can supply a clear understanding from the solar system to the entire universe on the basis of ground-based and space-based observations. Nowadays, with the improvement of the accuracy of astrometry, besides direct measurements of stellar positions and velocities, the indirect measurements can be extensively conducted to provide vital implications for the frontiers themes of habitable planets, black holes and dark matter, etc. SOFA service integrates the basic theory of astrometry and is a convenient tool to conduct astrometric related calculations and can be applied to a diverse research fields. This python package was originally intended to facilitate the use of SOFA services for various astronomical calculations. Thanks to the excellent, comprehensive features of SOFA services, this package can also be further used for the related astronomical calculation in python.
	
\begin{acknowledgements}
This work is financially supported by the National Natural Science Foundation of China (Grant Nos. 12033010, 11773081, 12111530175), the Strategic Priority Research Program on Space Science of the Chinese Academy of Sciences (Grant No. XDA 15020800), Foundation of Minor Planets of the Purple Mountain Observatory.

\end{acknowledgements}

\bibliographystyle{raa}
\bibliography{ms}
	
\label{lastpage}
	
\end{document}